\documentclass{amsart}   

\usepackage[applemac]{inputenc}
\usepackage{amsxtra}
\usepackage{txfonts}
\usepackage{mathrsfs}
\usepackage{braket}

\newcommand{\R}{\mathbb R}
\newcommand{\C}{\mathbb C}
\newcommand{\N}{\mathbb N}
\renewcommand{\H}{\mathcal H}
\newcommand{\E}{\mathbb E}
\newcommand{\eps}{\varepsilon}
\renewcommand{\mod}[1]{\lvert #1 \rvert}
\newcommand{\norm}[1]{\| #1 \|}
\newcommand{\Norm}[1]{\left\| #1 \right\|}
\newcommand{\I}{\text{Id}}
\newcommand{\J}{\text{i}}
\newcommand{\e}{\text{e}}

\newcommand{\hypothesis}[2]{
	\begin{itemize}
		\item[\textrm{(#1)}]\quad #2 
	\end{itemize}
}

\newcommand{\tr}{\mathop{\text{Tr}}}

\newcommand{\re}{\mathop{\text{Re}}}

\newcommand{\dom}{\mathop{\text{dom}}}
\newcommand{\spec}{\mathop{\text{spec}}}
\newcommand{\slim}{\mathop{\text{s-lim}}}

\theoremstyle{definition}
\newtheorem{definition}{Definition}[section]

\theoremstyle{plain}
\newtheorem{theorem}{Theorem}[section]
\newtheorem{proposition}[theorem]{Proposition}
\newtheorem{lemma}[theorem]{Lemma}
\newtheorem{corollary}[theorem]{Corollary}

\theoremstyle{remark}
\newtheorem{remark}[definition]{Remark}

\begin{document}

\title{Repeated Interaction Quantum Systems: van Hove Limits and Asymptotic States}

\author{Rodrigo Vargas}
\address{Institut Fourier, 
100 rue des Maths, BP 74, 38402 St Martin d'Hères, France.}
\email{rodrigo.vargas@ujf-grenoble.fr}

\begin{abstract}
We establish the existence of two weak coupling regime effective dynamics for an open quantum system of repeated interactions
(vanishing strength and individual interaction duration, respectively).
This generalizes known results \cite{AJ} in that the von Neumann algebras describing the system and the chain element may not be of finite type. Then (but now assuming that the small system is of finite type), we prove that both effective dynamics capture the long-term behavior of the system: existence of a unique asymptotic state for them implies the same property for the respective exact dynamics---provided that the perturbation parameter is sufficiently small. The zero-th order term in a power series expansion in the perturbation parameter of such an asymptotic state is given by the asymptotic state of the effective dynamics. We conclude by working out the case in which the small system and the chain element are spins.
\end{abstract}


\thanks{
This work was partially funded by Nucleus Millennium Information and Randomness P04-069-F
}
\dedicatory{Dedicated to Mariana Huerta}
\date{\today}

\maketitle

\section{Introduction}
\label{intro}

Recall that an open quantum system consists of a so-called \emph{small system} $\mathscr S$ immersed in a  reservoir $\mathscr R$, and that one is usually interested (perhaps by necessity) only in the observables of $\mathscr S$. In the repeated interaction model one assumes that the reservoir is an infinite chain of identical subsystems $\{\mathscr E_n\}_{n\in\N}$, called \emph{chain elements}, which interact with $\mathscr S$ sequentially, one at a time, in the order given by their labels $n\in\N$. Here we will suppose that:
\begin{itemize}
\item
	The time that $\mathscr S$ spends interacting with each $\mathscr E_n$---which could depend on $n$ or even be random---is actually constant, equal to $\tau>0$.
\item
	The way in which $\mathscr S$ interacts with each $\mathscr E_n$ is also independent of $n$.
\item
	All chain elements are initially in the same state.
\end{itemize}
More general models can be considered, as in \cite{BJM3,BJM2}.

Repeated interaction systems (RISs) have been used in connection with several domains, including quantum optics \cite{orszag} (in particular, regarding quantum state preparation \cite{WBKM}) and quantum noises \cite{A:qn,AP,Barchielli:cm}.
From an open systems point of view, they are interesting because of their mixture of sim\-pli\-ci\-ty---they have, by construction, a markovian nature---and thermodynamical non-triviality. {Sin\-ce} not much is known about statistical physics far from equilibrium, that makes them a promising source of examples and inspiration; nevertheless, their rigorous study is just in its beginnings. In this article, we focus on their perturbative analysis: we address the question of existence of van Hove effective dynamics and its use in studying the eventual asymptotic states, as we explain in what follows.

To place things in context, let us recall some known results about open systems with time-independent hamiltonian. In general, the evolution restricted to the small system satisfies a complicated integro-differential equation, and one is interested in finding asymptotic regimes in which the resulting effective dynamics is simpler. One possibility is to assume that the coupling between the small system and its environment is small, in which case one must rescale  time  so as to see the effects of the interaction: the dynamics is, then, composed of a fast part coming from the free evolution, and a slow part coming from the interaction. As it turns out, those dynamics decouple in the limit: the slow part, called van Hove limit, becomes markovian; the fast one becomes noise, which is the reason why the weak coupling regime is also called \emph{stochastic limit}~\cite{acc-lu-vol}. The mathematical study of the van Hove limit  was begun by Davies \cite{davies_markovian1} in 1974. The fact that the slow dynamics exists (at least in some cases) can be seen as one justification for the use of master equations when studying open systems.  The procedure which gives the generator of the effective dynamics can be understood as a dynamical Fermi golden rule; see \cite{DF:fgr} for an exposition of the subject. 
An interesting, somewhat unexpected result is the following: if the original system has an asymptotic state, then it is well approximated by the asymptotic state of its van Hove limit. Additional information on the subject can be found in \cite{Leb-Sp}.

The study of weak coupling regimes in the case of RISs was begun by Attal and Joye \cite{AJ}. As we will see later, there are at least two such regimes in this context: calling $\lambda$ the strength of the interacion, one has the cases $\lambda\rightarrow 0$, and $\tau\rightarrow 0$ as $\lambda^2\tau\rightarrow 0$. In \cite{AJ},  the existence of the slow dynamics is established for both regimes,  under the hypothesis that both the small system and the chain element are finite-dimensional. They also study a third regime ($\tau\rightarrow 0$ while $\lambda^2\tau$ is kept constant) which is not perturbative anymore; it has the interesting feature that one can always adjust the model in such a way that the effective dynamics is generated by any prescribed Lindbladian.

Our objective in this article is two-fold:
\begin{itemize} 
\item
	To generalise the results in \cite{AJ} to the infinite-dimensional case.
\item
	To study the extent to which the previously described relation between asymptotic states of a given system and its van Hove limit holds for RISs.
\end{itemize}
The precise meaning of \emph{asymptotic state} in this context is provided by
 Bruneau, Joye and Merkli \cite{BJM} who have proved, assuming that the small system is finite-dimensional and under an ergodicity hypothesis, that any given initial state of the small system converges, when $t\rightarrow\infty$, towards a unique time-periodic state. It is to be noted that this is not a state of thermal equilibrium, to start with because it is not constant, but above all because it has a non-vanishing entropy production; this justifies the claim above about the thermodynamical non-triviality of RISs.

\section{Mathematical setup}

Let $M_S$ and $M_E$ be two von Neumann algebras, meant to describe the small system and one individual chain element. Let $\alpha_S^t:M_S\rightarrow M_S$ and $\alpha_E^t:M_E\rightarrow M_E$ be the $*$-weakly-continuous groups of automorphisms which correspond to their free evolutions. We will suppose that $M_S$ and $M_E$ are mutually commuting subalgebras of a larger von Neumann algebra $M$ which is generated by them.\footnote{This amounts to identifying $M_S\cong M_S\otimes 1_E$, $M_E\cong 1_S\otimes M_E$ and letting $M=M_S\otimes M_E$.}
This permits to extend $\alpha_S^t,\alpha_E^t:M\rightarrow M$; we denote the derivations which generate these extended groups by $\delta_S$ and $\delta_E$, respectively, and we denote $\alpha_S^t\alpha_E^t$ simply by $\alpha_{SE}^t$. We write $\mathcal E_S$ and $\mathcal E_E$ for the set of normal states of $M_S$ and $M_E$, respectively.

Given a self-adjoint element $v\in M$, consider the perturbed dynamics $\varphi_{SE}^t$ generated by the derivation $\delta_S + \delta_E +\J\lambda[v,\cdot]$. It is explicitely given by the convergent  series
\begin{equation} \label{phiSE}
\varphi_{SE}^t = \bigl\{ \I_M + \sum_{k\geq 1} (\J\lambda)^k\varphi_{SE,k}^t \bigr\}\alpha_{SE}^t,
\end{equation}
where the $\varphi_{SE,k}^t$ are given by the $*$-weakly-convergent integrals
\begin{equation} \label{phiSEk}
\varphi_{SE,k}^t = \int_0^t dt_k \cdots \int_0^{t_2} dt_1\  \alpha_{SE}^{t_1}[v,\cdot]\alpha_{SE}^{-t_1} \cdots \alpha_{SE}^{t_k}[v,\cdot]\alpha_{SE}^{-t_k}.
\end{equation}
We are interested in the repeated interaction evolution restricted to the small system, under the assumption that all chain elements are initially in the $\beta$-KMS state $\omega_E\in \mathcal E_E$. Therefore, we consider
\begin{equation} \label{phiRes}
\varphi_\text{res}^t = \bigl(\E_S \varphi_{SE}^\tau\bigr)^n \E_S\bigl.\varphi_{SE}^{t_1}\bigr|_{M_S} : M_S\rightarrow M_S,
\end{equation}
where $n\in\N$, $t_1\in \left[0,\tau\right[$, $t=n\tau+t_1$ and $\E_S:M\rightarrow M_S$ is the conditional expectation given by
\begin{equation} \label{ES}
\E_S(x_Sx_E) = x_S \omega_E(x_E),\quad\forall x_S\in M_S, x_E\in M_E.
\end{equation}
\begin{remark}
The existence of $\E_S$ follows from the fact that, under the isomorphisms $M_S\cong M_S\otimes 1_E$, $M_E\cong 1_S\otimes M_E$ and $M\cong M_S\otimes M_E$, it can be written as the composition
\[
M_S\otimes M_E \xrightarrow{\I_{M_S}\otimes\omega_E} M_S\otimes \C \cong M_S \xrightarrow{(\cdot)\otimes 1_E} M_S\otimes M_E.
\]
\end{remark}

Equation \eqref{phiRes} defines a $*$-weakly continuous family of completely positive maps. Observe that the semigroup property fails, since $\varphi_\text{res}^t$ gives the correct time evolution only if we start at times which are integer multiples of $\tau$. Note, however, that one can define in the obvious way a two-parameter family $\{\varphi_\text{res}^{t_1,t_2}\}_{t_1\leq t_2\in\R}$
satisfying $\varphi_\text{res}^t = \varphi_\text{res}^{0,t}$ and $\varphi_\text{res}^{t_1,t_2}\varphi_\text{res}^{t_2,t_3} = \varphi_\text{res}^{t_1,t_3}$, for all $t_1\leq t_2\leq t_3$.
This is also related to the fact that our intuitively correct formula for time evolution can be obtained by exponentiation of a time-dependent hamiltonian---which would be somewhat more rigorous. In fact, one could consider the von Neumann algebra which describes simultaneously the small system and the \emph{entire} chain, and define there a hamiltonian which, depending on the instant of time, makes the small system interact with the adequate chain element. One would obtain a piecewise constant generator whose exponentiation, after composition with the right conditional expectation projecting onto the small system, coincides with $\varphi_\text{res}^t$. We will omit the simple but lengthy and notationally involved proof of this fact, because it does not give any insight on the problems which concern us in this work. For more details, see \cite{AJ,BJM}.

To simplify the study of the weak coupling regime, we will impose a condition on the perturbation $v\in M$ which ensures that there are no first order effects:
\hypothesis{H1}{
	There exists a projection  $p_0\in M_E$,  invariant under  $\alpha_E^t$,  such that
	\[v = p_0v(1-p_0) + (1-p_0)vp_0.\]
}
\begin{remark}
First order effects (as can be seen from the Dyson series) do not reflect an influence from the environment: they come from the part of the perturbation which can be interpreted as modifying the free dynamics of the small system. 
\end{remark}
One can think of  $p_0$ as the projection onto the first eigenspace of $\delta_E$, which could be interpreted as an \emph{absolute vacuum} state. In this case, Hypothesis (H1) is loosely saying that the small system and the chain element interact only through creation and annihilation processes at the chain level. To see this in more detail, we refer the reader to \cite{AJ}, where  an interaction which precisely falls within this description is considered. But Hypothesis (H1) can perfectly apply in other, different situations, where the interpretation just given is not adequate. We should warn, however, against one potentially tempting interpretation: by GNS construction we can always assume that $\omega_E(x) = \langle \Omega_E,x\Omega_E\rangle$, with $\Omega_E$ belonging to a Hilbert space on which $M_E$ acts. The projection $\ket{\Omega_E}\bra{\Omega_E}$ \emph{cannot} take the role of $p_0$ because it does not belong to $M_E$.

\begin{proposition} \label{prop_T}
The linear operator
\begin{equation} \label{T}
T(\lambda,\tau) = \E_S\bigl.\varphi_{SE}^\tau\bigr|_{M_S} \in B(M_S)
\end{equation}
is completely positive, normal  and $\norm{T(\lambda,\tau)} = 1$. Moreover, given $\tau>0$, the map $\lambda\mapsto T(\lambda,\tau)$ is analytic and, if the hypothesis (H1) holds, it is also even.
\end{proposition}
\begin{proof}
The convergence of the Dyson series shows that $\lambda\in\R\mapsto \varphi_{SE}^\tau \in B(M)$ is analytic; it follows that $\lambda\mapsto T(\lambda,\tau)$ is analytic too, since
\[
F\in B(M) \mapsto \E_S F|_{M_S} \in B(M_S)
\]
is linear and bounded (observe that $\E_S$, being a conditional expectation, has norm 1). Complete positivity and normality are a consequence of the fact that $\E_S$ and $\varphi_{SE}^t$ have these properties. Since $T(\lambda,\tau)1 = 1$, by general properties of completely positive maps we also have that $\norm{T(\lambda,\tau)} = 1$.

Let us check the parity. Under the hypothesis (H1), the invariance of $p_0$ under the free evolution $\alpha_E^t$ implies---thanks to the KMS condition---that
\begin{align*}
\E_S(p_0x_Sx_E) &=
\omega_E(p_0x_E)x_S \\ &= \omega_E\bigl(x_E\alpha_E^{\J\beta} (p_0)\bigr)x_S \\ &= \omega_E(x_Ep_0)x_S \\
	&= \E_S(x_Sx_Ep_0),\quad \forall x_S\in M_S,x_E\in M_E. 
\end{align*}
Hence,
\[
\E_S(x) = \E_S(p_0xp_0) + \E_S((1 - p_0)x(1 - p_0)), \quad\forall x\in M.
\]
Using this, all we have to do is prove that, for all odd $k$ and $x_S\in M_S$, \[
p_0\varphi_{SE,k}^\tau(x_S) p_0 = (1-p_0) \varphi_{SE,k}^\tau(x_S) (1-p_0) = 0,
\]
where $\varphi_{SE,k}^t$ is defined in \eqref{phiSEk}.
But this follows again from the invariance of $p_0$ and the relations
\begin{align*}
p_0[v,x_S] &= p_0vx_S(1-p_0) - p_0x_Sv(1-p_0) \\ &= [v,x_S](1-p_0), \\
[v,x_S]p_0 &= (1-p_0)[v,x_S],
\end{align*} 
which are a consequence of the fact that $p_0$ and $x_S$ commute.
\end{proof}

\section{Van Hove limit}

Schematically, we are concerned with the  study of an operator of the form
\begin{equation} \label{aproximation}
(P\e^{\tau(A+\lambda B)}P)^n \approx \bigl[ (1+\tau O(\lambda^2\tau)\e^{\tau A}) \bigr]^n,
\end{equation}
where $P$ is a projection, $A$ the generator of a group of isometries, $B$ a perturbation and $n\in\N$. Note that the parameter that determines the perturbative nature of a given regime is $\lambda^2\tau$; thus, we can immediately identify three different perturbative regimes:
\begin{enumerate}
\item 
	$\tau$ is kept constant, in which case $\lambda$ must go to zero.
\item
	$\tau\rightarrow 0$. Now, $\lambda$ can go to zero, remain bounded or even diverge---provided $\lambda^2\tau\rightarrow 0$.
\item
	$\tau\rightarrow\infty$ and $\lambda^2\tau\rightarrow 0$.
\end{enumerate}
In this article we treat the first two cases. The third one, which is a priori out of the reach of our method,  seems to oscilate with $\tau$ (the example of Section \ref{example} gives some evidence of this).

To identify the adequate time scale of an effective dynamics in each of these regimes, note that the approximation \eqref{aproximation} is likely to become useless when $n \approx 1/(\lambda^2\tau^2)$---that is, when $t=n\tau \approx 1/(\lambda^2\tau)$.
Therefore, the appropriate time scale should be $s=\lambda^2\tau t$, irrespective of the perturbative regime which is being considered.

\subsection{A preliminary result}

Here we state a simple generalization of a theorem by Davies \cite{davies}, which is an abstract weak coupling dynamics existence result. 

\begin{theorem} \label{davies_infini}
Let $X$ be a Banach space, $A_0:\dom A_0\subset X\rightarrow X$ the generator of a strongly continuous  group  of isometries and $A_1: \R\rightarrow B(X)$ a norm-continuous map. Suppose that
\[
\underset{T\rightarrow\infty}{\slim} \frac{1}{T}\int_0^T dt\ \e^{tA_0}A_1(0)\e^{-tA_0}
\]
exists and denote it by $A_1(0)^\natural$. Then, defining $A(\eps) = A_0 + \eps A_1(\eps)$, we have that
\[
\lim_{\eps\rightarrow 0} \sup_{s\in [0,s_0]} \Norm{ \bigl(\e^{sA(\eps)/\eps}\e^{-sA_0/\eps} - \e^{sA_1(0)^\natural}\bigr)x } = 0,
\]
for any $s_0>0$ and $x\in X$.
\end{theorem}
\begin{proof}
Davies proved this result when $A_1(\eps)$ is actually constant; we will get the general case as a consequence, by showing that
\[
\lim_{\eps\rightarrow 0} \sup_{s\in [0,s_0]} \Norm{\e^{sA(\eps)/\eps} -\e^{s(A_0+\eps A_1(0))/\eps} } = 0
\]
and using the triangle inequality.
By Duhamel's formula,
\begin{align*}
&\e^{sA(\eps)/\eps} -\e^{s(A_0+\eps A_1(0))/\eps} \\ &\quad = \int_0^s ds_1\ \e^{s_1 A(\eps)/\eps} (A_1(\eps)-A_1(0))\e^{(s-s_1)(A_0+\eps A_1(0))/\eps}.
\end{align*}
Now, apply the Dyson expansion and use Remark \ref{err_dyson} to get the estimate
\[ 
\Norm{\e^{s_1A(\eps)/\eps} } = \Norm{\e^{s_1(A_0/\eps + A_1(\eps)) }} 
	\leq 1 + \sum_{k\geq 1} \frac{s_1^k}{k!} \norm{A_1(\eps)}^k,
\] 
which by continuity is bounded uniformly in $\eps$. Similar considerations apply to
\[
\norm{\e^{(s-s_1)(A_0+\eps A_1(0))/\eps} },
\]
from which the claim follows.
\end{proof}

\begin{remark} \label{rem_davies_infini}
The strong limit  
\[
A_1(0)^\natural = \underset{T\rightarrow\infty}\slim \frac{1}{T}\int_0^Tdt\ \e^{tA_0}A_1(0)\e^{-tA_0}
\] 
is the so-called \emph{spectral averaging} of $A_1(0)$ with respect to the spectrum of $A_0$. 
There are at least two known conditions which ensure its existence \cite{DF:fgr}, namely:
\begin{enumerate}
\item
	$A_0$ admits a total set of eigenvectors, and
\item
	$A_1(0)$ is compact and $X$ is a Hilbert space.
\end{enumerate}
In the first case, $A_1(0)^\natural$ is equal to
\[
\sum_n P_n A_1(0) P_n,
\]
where the $P_n$'s are the spectral projections of $A_0$ and the sum converges strongly. Observe that it is, in a sense, the part of $A_1(0)$ which commutes with $A_0$---and this interpretation holds whenever the strong limit $A_1(0)^\natural$ exists.
\end{remark}

\subsection{The regime $\lambda\rightarrow 0$}

To use Theorem \ref{davies_infini} in the repeated interaction case we start by restricting our attention to the discrete semigroup consisting of integer powers of $T(\lambda,\tau)$; otherwise said, we regard only times which are integer multiples of $\tau$. The only problem then is to ``interpolate'' the semigroup $\{T(\lambda,\tau)^n\}_{n\in\N}$ to continuous time.

\begin{theorem} \label{lambda1}
Suppose that Hypothesis (H1) holds, as well as 
\hypothesis{H2a}{
	The spectrum of  $\alpha_S^\tau$  is not dense in the circle  $S^1\subseteq \C$.
}
Let $\Gamma\subseteq \C$ be a curve with $\deg(\Gamma,0)=0$ which encircles the spectrum of $\alpha_S^\tau$, choose a branch of logarithm analytic in the interior of $\Gamma$, and define 
\[
A_0 = \frac{1}{2\pi \J} \int_{\Gamma} dz\ \log (z)(z-\alpha_S^\tau)^{-1}.
\]
Assume, finally, that
\hypothesis{H3a}{
	\( \displaystyle
	(\E_S\varphi_{SE,2}^\tau)^\natural = \underset{T\rightarrow\infty}\slim \frac{1}{T}\int_0^Tdt\ \e^{tA_0} \E_S\varphi_{SE,2}^\tau\e^{-tA_0}
	\)
	exists.
}
Then, the norm-continuous contraction semigroup 
\[
\varphi_\text{eff}^s =\e^{-s(\E_S\varphi_{SE,2}^\tau)^\natural}:M_S\rightarrow M_S
\] 
satisfies
\[
\lim_{\lambda\rightarrow 0} \sup_{s\in[0,s_0]} \Norm{ \bigl( T(\lambda,\tau)^{\lfloor s/(\lambda^2\tau) \rfloor} \alpha_S^{-\tau\lfloor s/(\lambda^2\tau) \rfloor} - \varphi_\text{eff}^s\bigr) x } = 0, 
\]
for all $s_0>0$. Here, $\lfloor\cdot\rfloor$ denotes the integer part of its argument.
\end{theorem}
\begin{proof}
Recall that $T(\lambda,\tau) = {\E_S\varphi_{SE}^\tau}\bigr|_{M_S}$, whence $T(0,\tau) = \alpha_S^\tau$ and there exists an $\eps>0$ such that the curve $\Gamma$ encircles the spectrum of $T(\lambda,\tau)$, for all $\lambda^2<\eps$. Define $A:\left]-\eps,\eps\right[ \rightarrow B(M)$ by
\[
A(\lambda^2) = \frac{1}{2\pi \J} \int_{\Gamma} dz\ \log (z)(z-T(\lambda,\tau))^{-1},
\]
which gives an analytic function since the dependence of $T$ in $\lambda$ is quadratic. Assuming that
\[
A'(0)^\natural = \underset{T\rightarrow \infty}\slim \frac{1}{T} \int_0^T dt\ \e^{tA_0}A'(0)\e^{-tA_0}
\]
exists, Theorem \ref{davies_infini} would provide the conclusion with $\varphi_\text{eff}^s =\e^{sA'(0)^\natural}$. Therefore, we have to prove that $A'(0)^\natural$ exists and is equal to $-(\E_S\varphi_{SE,2}^\tau)^\natural$.
To do that,
recall that 
\begin{align*}
T(\lambda,\tau) &= \alpha_S^\tau - \lambda^2\E_S\varphi_{SE,2}^\tau\alpha_S^\tau + O(\lambda^4) \\ &=: \alpha_S^\tau + \lambda^2 T_2 + O(\lambda^4).
\end{align*}
Hence, 
\begin{align*}
A'(0)^\natural &= 
\underset{T\rightarrow\infty}\slim \frac{1}{T}\int_0^Tdt \frac{1}{2\pi \J} \int_{\Gamma} dz\  \log(z) (z-\alpha_S^\tau)^{-1} \e^{tA_0} T_2\e^{-tA_0} (z-\alpha_S^\tau)^{-1}  
\\
&= \frac{1}{2\pi \J} \int_\Gamma dz\ \log(z) (z-\alpha_S^\tau)^{-1}  \biggl(\underset{T\rightarrow\infty}\slim \frac{1}{T}\int_0^T dt\ \e^{tA_0} T_2\e^{tA_0} \biggr) (z-\alpha_S^\tau)^{-1} 
\\
&=  T_2^\natural\frac{1}{2\pi \J} \int_\Gamma dz\ \log(z)(z-\alpha_S^\tau)^{-2} \\ &= T_2^\natural \frac{d}{dz}\biggr|_{z=\alpha_S^\tau} \log(z) \\
&= (-\E_S\varphi_{SE,2}^\tau\alpha_S^\tau)^\natural\alpha_S^{-\tau} = -(\E_S\varphi_{SE,2}^\tau)^\natural.
\end{align*}
The integration order can be reversed, since the integrand and the domain of integration are both bounded; the same argument justifies the exchange of strong limit and complex integral. Note that Hypothesis (H3a) ensures the existence of the limit.
\end{proof}

\begin{remark}
The spectral projections of $A_0$ (which is always \emph{bounded}) do not necessarily coincide with those of $\delta_S$, so that $\varphi_\text{eff}^s$ and $\alpha_S^t$ do not necessarily commute. 
An extreme case of this would be a harmonic oscilator with energy spectrum $\{2\pi n/\tau: n\in\N\}$. Then, if we take $\log r\e^{\J t} = \log r +\J t$ with $t\in \left] -\pi,\pi \right[$, we get $A_0 = 0$.
\end{remark}

\begin{remark} \label{rk_lambda}
In \cite{AJ}, Attal and Joye prove Theorem \ref{lambda1} when the Hilbert spaces $\H_S$ and $\H_E$ upon which $M_S$ and $M_E$ act, respectively, are finite dimensional. Their method consists in solving explicitely the equation
\begin{equation} \label{AJ}
T(\lambda,\tau) =\e^{\tau(A_0 + \lambda^2 A_1)} + O(\lambda^4),
\end{equation}
where $A_0$ and $A_1$ are the unknowns. Our method, although conceptually simpler, is essentially the same. Note that the use of a logarithm makes things easier but does not provide an optimal result, since in infinite dimension it might be possible that equation \eqref{AJ} admits a solution, even if the spectrum of $\alpha_S^\tau$ is dense in the unit circle.
\end{remark}

Theorem \ref{lambda1} actually allows one to understand the behavior of $\varphi^t_\text{res}$ for $\lambda\ll 1$ and arbitrary $t\lesssim 1/\lambda^2$; in other words, the restriction to times which are integer multiples of $\tau$ is immaterial. 
\begin{corollary} \label{lambda2}
Under the same hypothesis of Theorem \ref{lambda1}, the contraction semi-group $\varphi_\text{eff}^s:M_S\rightarrow M_S$ satisfies also
\[
\lim_{\lambda\rightarrow 0} \sup_{s\in[0,s_0]} \Norm{ \bigl( \varphi_\text{res}^{s/\lambda^2}\alpha_S^{-s/\lambda^2} - \varphi_\text{eff}^s \bigr)x } = 0,\quad \forall s_0>0.
\]
\end{corollary}
\begin{proof}
Indeed, writing $s/\lambda^2 = n\tau+t_1$ with $n=\lfloor s/(\lambda^2\tau) \rfloor$, one has
\begin{align*}
\left\| \varphi_\text{res}^{s/\lambda^2}\alpha_S^{-s/\lambda^2} - \varphi_\text{eff}^s \right\| 
	&= \Norm{ T(\lambda,\tau)^n\E_S\varphi_{SE}^{t_1}\alpha_S^{-(n\tau +t_1)} - \varphi_\text{eff}^s } \\
	&\leq \Norm{ T(\lambda,\tau)^n\alpha_S^{-n\tau} - \varphi_\text{eff}^s }  + \Norm{ T(\lambda,\tau)^n\E_S\bigl( \varphi_{SE}^{t_1}\alpha_S^{-t_1} - \I_M \bigr)\alpha_S^{-n\tau} }.
\end{align*}
The first term is controlled by Theorem \ref{lambda1}, while, using the Dyson expansion, Remark \ref{err_dyson} and the fact that $\E_S\varphi_{SE,1}^{t_1} = 0$, the second is bounded by
\[
\Norm{ \E_S(\varphi_{SE}^{t_1} - \alpha_{SE}^{t_1}) } \leq f_2(\lambda^2 \tau)\lambda^2 \tau.
\]
\end{proof}

\subsection{The regime $\tau\rightarrow 0$, $\lambda^2\tau\rightarrow 0$}

This regime is, analitically, somewhat more delicate, because one has to control the dependence in $\tau$ of the error as $\lambda^2\tau\rightarrow 0$. That prevents us from just using functional calculus as in the previous subsection. In \cite{AJ}, Attal and Joye use a refined, but finite dimensional, version of Theorem \ref{davies_infini} to deal with this; however, their proof cannot be easily extended to the infinite dimensional case.  We take a different approach, which consists essentially in regrouping the error terms so that one can apply Theorem \ref{davies_infini} directly. 


\begin{lemma}  \label{lem_tau}
Given constants $\lambda_0,\tau_0>0$, we say that $\lambda$ and $\tau$ are \emph{admissible} if
\[
\tau \in [0,\tau_0],\quad \lambda^2\tau\in [0,\lambda_0^2\tau_0].
\]
Suppose that there exists  some $C_0>0$ such that, for all admissible $\lambda$ and $\tau$,
\[
\bigl\| T(\lambda,\tau) - \{1 + \lambda^2\tau^2 A_1\} \alpha_S^\tau \bigr\| \leq C_0\lambda^2\tau^3,
\] 
where $A_1\in B(M_S)$ is such that
\(
t\mapsto \alpha_S^tA_1\alpha_S^{-t}
\)
is norm continuous. Then, again for all admissible $\lambda$ and $\tau$,
\[
\sup_{0\leq s\leq s_0} \bigl\| \varphi_{\text{res}}^{s/(\lambda^2\tau)} -\e^{s(\delta_S + \lambda^2\tau A_1)/(\lambda^2\tau)} \bigr\| 
= O(\tau),
\]
where $s_0>0$ is arbitrary. 
\end{lemma}
\begin{proof}
Thanks to the Dyson series, with $\eps=\lambda^2\tau$ in the notation of Appendix \ref{appDyson},
\begin{align*}
T(\lambda,\tau) &-\e^{\tau(\delta_S + \lambda^2\tau A_1)} \\ &=  \biggl\{ \lambda^2\tau^2 A_1 - \lambda^2\tau \int_0^{\tau} dt\ \alpha_S^{t} A_1 \alpha_S^{-t}\biggr\} \alpha_S^{\tau} + E(\lambda,\tau), \end{align*}
where, using the function $f_2$ defined in \eqref{f_n},
\[ 
\Norm{E(\lambda,\tau)} \leq C_0\lambda^2\tau^3 + f_2(\lambda^2\tau^2)(\lambda^2\tau)^2\tau^2 = O(\lambda^2\tau^3).
\] 
Moreover, by continuity of $t\mapsto \alpha_S^{t}A_1\alpha_S^{-t}$, 
\[
\int_0^\tau dt\ \alpha_S^{t} A_1 \alpha_S^{-t} = \tau A_1 + O(\tau^2)
\] 
and we conclude that, for all admissible $\lambda$ and $\tau$,
\[
\Norm{T(\lambda,\tau) -\e^{\tau(\delta_S + \lambda^2\tau A_1)}} \leq C_1\lambda^2\tau^3,
\] 
where the constant $C_1$ depends only on $C_0$, $\lambda_0$, $\tau_0$ and $A_1$. Now, a standard telescope expansion shows that
\begin{align*}
\biggl\| T(\lambda,\tau)^m - \bigl[\e^{\tau(\delta_S+\lambda^2\tau A_1)} \bigr]^m \biggr\| &= \biggl\| \sum_{k=1}^m T(\lambda,\tau)^{k-1} \bigl( T(\lambda,\tau) -\e^{\tau(\delta_S+\lambda^2\tau A_1)} \bigr) \bigl[\e^{\tau(\delta_S+\lambda^2\tau A_1)}\bigr]^{m-k} \biggr\| \\
	&\leq m \Norm{\e^{\tau(\delta_S+\lambda^2\tau A_1)}} ^m O(\lambda^2\tau^3).
\end{align*}
But we also have, this time using the Dyson series with $\eps=\lambda^2\tau^2$, that 
\[
\Norm{\e^{\tau(\delta_S+\lambda^2\tau A_1)}} \leq 1 + C_2\lambda^2\tau^2,\quad C_2=f_1(\lambda^2\tau^2),
\]
whence
\begin{align*}
\biggl\| T(\lambda,\tau)^m - \bigl[\e^{\tau(\delta_S+\lambda^2\tau A_1)} \bigr]^m \biggr\| 
	&\leq C_1s_0\tau \bigl( 1+C_2\lambda^2\tau^2 \bigr)^{s/(\lambda\tau)^2} \\
	&\leq C_1s_0\tau\e^{s_0\log( 1+C_2\lambda^2\tau^2 )/(\lambda\tau)^2} = O(\tau).
\end{align*}
We conclude by observing that, writing $s/(\lambda^2\tau) = m\tau+t_1$ with $m=\lfloor s/(\lambda\tau)^2 \rfloor$,
\begin{align*}
\biggl\| \varphi_{\text{res}}^{s/(\lambda^2\tau)} &- \bigl[\e^{\tau(\delta_S+\lambda^2\tau A_1)} \bigr]^{s/(\lambda\tau)^2} \biggr\| \\
	&\leq \Norm{ T(\lambda,\tau)^m \bigl( \alpha_S^{t_1} -\e^{t_1(\delta_S+\lambda^2\tau A_1)} \bigr)} + \Norm{ \bigl( T(\lambda,\tau)^m -\e^{m\tau( \delta_S + \lambda^2\tau A_1 )} \bigr)\e^{t_1(\delta_S+\lambda^2\tau A_1)}}\\
	&\leq f_1(\lambda^2\tau^2)\lambda^2\tau^2 + O(\tau) \bigl( 1+f_1(\lambda^2\tau^2)\lambda^2\tau^2 \bigr),
\end{align*} 
which is of $O(\tau)$ for all admissible $\lambda$ and $\tau$.
\end{proof}

\begin{theorem} \label{tau}
Suppose that Hypothesis (H1) holds,  as well as 
\hypothesis{H2b}{
	\( \displaystyle
	(\E_S[v,\cdot]^2)^\natural = \underset{T\rightarrow\infty}\slim \frac{1}{T}\int_0^T dt\ \alpha_S^t \E_S[v,\cdot]^2 \alpha_S^{-t}
	\)
	exists,
}
\hypothesis{H3b}{
	\( \displaystyle 
	t\in\R \mapsto \alpha_{SE}^{t}[v,\cdot]\alpha_{SE}^{-t}\in B(M)
	\)
	is norm continuous.
}
Let $\tau_n,\lambda_n\geq 0$ be two sequences such that $\tau_n\rightarrow 0, \lambda_n^2\tau_n\rightarrow 0$. Then, the semigroup $\varphi_\text{eff}^s =\e^{-\frac{s}{2}(\E_S[v,\cdot]^2)^\natural}:M_S\rightarrow M_S$ satisfies
\[
\lim_{n\rightarrow\infty} \sup_{\substack{ 0\leq s\leq s_0 \\ \norm{x}\leq 1 }} \bigl(\rho, \bigl(  \varphi_\text{res}^{s/(\lambda_n^2\tau_n)} \alpha_S^{-s/(\lambda_n^2\tau_n)} - \varphi_\text{eff}^s \bigr) x \bigr)_{(M_S)_*,M_S} = 0,
\]
for all fixed $\rho\in (M_S)_*$ and $s_0>0$. 
\end{theorem}
\begin{proof}
Observe, first, that by continuity of $t \mapsto \alpha_{SE}^{-t}[v,\cdot]\alpha_{SE}^t$ one has
\begin{align*}
\varphi_{SE,2}^\tau &= \int_0^\tau dt_2 \int_0^{t_2} dt_1 \alpha_{SE}^{t_1} [v,\cdot] \alpha_{SE}^{t_2-t_1} [v,\cdot] \alpha_{SE}^{-t_2} \\
	&= \frac{\tau^2}{2} [v,\cdot]^2 + O(\tau^3),
\end{align*}
since linear operator composition $B(M)\times B(M) \rightarrow B(M)$ is norm-continuous. Therefore, using Dyson's expansion and the evenness of $T(\lambda,\tau)$ in $\lambda$, one finds that 
\begin{align*}
T(\lambda,\tau) &= \E_S\biggl\{ 1+(\J\lambda)^2\biggl(\frac{\tau^2}{2}[v,\cdot]^2 + O(\tau^3)\biggr) \biggr\}\alpha_{SE}^\tau + O(\lambda^4\tau^4) \\
	&= \biggl\{ 1 -\frac{\lambda^2\tau^2}{2} \E_S[v,\cdot]^2 \biggr\}\alpha_S^\tau + O(\lambda^2\tau^3),
\end{align*}
where we have used the fact that $\E_S\alpha_{SE}^t = \alpha_S^t\E_S$ and $O(\lambda^4\tau^4)$ is, actually, $O(\lambda^2\tau^3)$ when $\lambda\leq C\tau^{-1/2}$. To apply Lemma \ref{lem_tau} we have to check that
\[
t\mapsto \alpha_S^{t}\E_S[v,\cdot]^2\alpha_S^{-t} = \E_S\bigl(\alpha_{SE}^{t}[v,\cdot]\alpha_{SE}^{-t}\bigr)^2\E_S
\]
is continuous, which is direct by hypothesis.

To conclude we would like to use Theorem \ref{davies_infini}, but the group
\[
\e^{t(\delta_S-\frac{\lambda^2\tau}{2}\E_S[v,\cdot]^2)} : M_S\rightarrow M_S
\]
is only $*$-weakly-continuous; we have to show that it admits a predual, which then by definition would  be strongly continuous. But we know that $\delta_S$ admits a predual (the generator of the strongly continuous group $(\alpha_S^t)_*$), and therefore it suffices to see that $(\E_S[v,\cdot]^2)^*: M_S^*\rightarrow M_S^*$ leaves the sub-space of ultraweakly continuous forms invariant. Now, for that it is enough that $\E_S[v,\cdot]^2: M_S\rightarrow M_S$ be ultraweak-ultraweak continuous, and, since $\E_S$ is positive and normal, all we have to do is prove that the operations $M_S\rightarrow M$ of left and right multiplication by elements of $M$ are ultraweak-ultraweak continuous---which is an elementary property of the ultraweak topology, concluding the proof.
\end{proof}

\begin{remark}
When the small system and the chain element are finite-dimensional, the hypothesis on the continuity of $\alpha_{SE}^{t} [v,\cdot] \alpha_{SE}^{-t}$ always holds; hence, this theorem is a generalization of the one in \cite{AJ}.
\end{remark}

\section{Asymptotic state}

In this section we will suppose that the von Neumann algebra $M_S$ is of finite type~$\text{I}_n$---that is, isomorphic to $\text{M}_n(\C)$. Recall that in this case all semigroups are automatically norm-continuous.

The expression ``asymptotic state'' in the context of quantum dynamics presupposes that the system is being studied in the Schrödinger picture; if we actually have a completely positive semigroup $\varphi^t:M_S\rightarrow M_S$, the evolution of states is given by
\[
\omega_t(x) = \omega_0(\varphi^t(x)),\quad x\in M_S,\ \omega_0\in \mathcal E_S.
\]
Now, the convergence $\omega_t \xrightarrow[t\rightarrow\infty]{} \omega_\infty$ for every state $\omega_0$ implies the  weak convergence of $\varphi^t(x)$ towards a limit $P(x)$ which defines a linear function $P:M_S\rightarrow M_S$. Note that $P(x)$ must be a multiple of the identity, because otherwise $\omega_0(P(x))$ would depend on $\omega_0$; therefore, 
\[
\varphi^t\xrightarrow[t\rightarrow\infty]{}P,\quad P(x)=\omega_\infty(x)1.
\]
Conversely, the convergence of $\varphi^t$ to a rank-one projection $P:M_S\rightarrow M_S$ (whose range must be $\C 1\subseteq M_S$ since $\varphi^t(1)=1$) implies the existence of a unique asymptotic state.

In the case of repeated interaction systems, one must take into account the fact that the asymptotic state, if it exists, is, in general, $\tau$-periodic \cite{BJM}; an obvious necessary condition for its existence is, then, that $T(\lambda,\tau)^n\rightarrow P(\lambda)$. In the next subsection we study this situation from an abstract viewpoint.

\subsection{On the analytic perturbation theory of matrices}

In this subsection we will suppose that $T:\left]-\eps_0,\eps_0\right[\rightarrow \text{M}_n(\C)$ is an analytic function such that $1\in\spec T(\eps)$ and $\norm{T(\eps)}=1$.
The classical reference for this material is \cite{kato}. We start with a lemma which lies at the heart of the section.

\begin{lemma} \label{P(0+)}
For each $\eps\in\left]-\eps_0,\eps_0\right[$, let $P(\eps)$ be the spectral projection of $1\in\spec T(\eps)$. Suppose that $T(\eps)^n \xrightarrow[n\rightarrow\infty]{} P(\eps)$ when $0<\eps<\eps_0$. Then,
	\begin{enumerate}
	\item
		$0\in \spec P(0)T'(0)P(0)$. Let $Q$ be its spectral projection. 
	\item
		$P(0)Q = QP(0)$, so $P(0)Q$ is a projection, too.
	\item
		$P(0^+) = \lim_{\eps\searrow 0} P(\eps)$ exists and is a sub-projection of $P(0)Q$.
	\end{enumerate}
\end{lemma}
\begin{proof}
Let $\tilde T$ be an analytic extension of $T$ to a complex neighbourhood of zero $\Omega\subseteq \C$. We want to prove, in the terminology of \cite{kato}, that 0 is not a branch point of $1\in\spec \tilde T(0)$. Since exceptional points are isolated, in any case we can suppose that there exist $m$ analytic functions $\tilde P_i:\Omega\setminus \left]-\infty,0\right]\rightarrow \text{M}_n(\C)$, which are all spectral projections of $\tilde T$, such that
\[
P(0) = \sum_{i=1}^m \tilde P_i(0).
\]
Now, one of these spectral projections, say $\tilde P_1 =: \tilde P$, must correspond to the eigenvalue $1\in\spec T(\eps)$ and must therefore be an analytic extension of $P$. Suppose, by contradiction, that $0$ is a branch point of $1$ of order $p-1\geq 1$. We know (see \cite[Theorem 1.9]{kato}) that, in this case, $\tilde P$ admits a Laurent expansion in powers of $z^{1/p}$ which necessarily contains negative powers. However, by continuity of the norm, we have
\[
\norm{P(\eps)} = \lim_{n\rightarrow\infty} \norm{T(\eps)^n} \leq \liminf_{n\rightarrow\infty} \norm{T(\eps)}^n = 1.
\]
This means that, if we approach through the real positive axis, $\lim_{\eps\rightarrow 0^+}\norm{ \tilde P(\eps) } = 1$. This is a contradiction, and we conclude that $0$ is not a branch point of $1$. In particular,  $\tilde P$ can be further extended to an analytic continuation of $P|_{\left]0,\eps_0\right[}$ defined on a complex neighbourhood of 0 and $P(0^+)$ exists.

Making use of $\tilde P(z)$, each $\xi_0\in P(0^+)\C^n$ yields an analytic choice $\xi(z) = \tilde P(z)\xi_0$ of eigenvectors of $\tilde T(z)$ with eigenvalue 1. Now, the first order term in $z$ in the equation $\tilde T(z)\xi(z) = \xi(z)$ is
\[
T(0)\xi'(0) + T'(0)\xi_0 = \xi'(0)
\]
which, pre-multiplied by $P(0)$, gives $P(0) T'(0) P(0)\xi_0 = 0$. In particular, 
\[
0\in\spec P(0) T'(0) P(0).
\]
Let $Q$ be its spectral projection. This means that $Q\xi_0=\xi_0$ for all $\xi_0\in P(0^+)\C^n$, and therefore that $P(0^+) = QP(0^+)$.

Next, we show that  $P(0^+) = P(0^+)Q$. This follows from  applying the same reasoning above to the (real) analytic function $T(\eps)^*$. In fact: it satisfies the hypothesis of the lemma; the spectral projection of $1\in\spec T(\eps)^*$ is $P(\eps)^*$; and we have that
\[
\frac{d}{d\eps}\biggr|_{\eps=0} T(\eps)^* = T'(0)^*.
\]
Therefore, we conclude that $P(0^+)^* = Q^*P(0^+)^*$.

Finally, since $Q$ is obtained by spectral calculus from $P(0)T'(0)P(0)$ and 
\[
[P(0),P(0)T'(0)P(0)]=0,
\]
 we have $[Q,P(0)]=0$. To conclude, it only remains to show that 
\[
P(0)P(0^+) = P(0^+)P(0) = P(0^+),
\]
for in that case $P(0^+) = P(0^+)P(0)Q = P(0)QP(0^+)$. But again, the equation $P(0)P(0^+)=P(0^+)$ just amounts to saying that the elements in $P(0^+)\C^n$ are eigenvectors of $T(0)$ with eigenvalue 1, and $P(0^+)P(0)=P(0^+)$ follows from applying the same reasoning to $T(\eps)^*$.
\end{proof}
\begin{remark}
Note that, by analyticity of $\tilde P$, one has $P(\eps) = P(0^+) + O(\eps)$ for $\eps>0$.
\end{remark}

The next result, which has some independent interest, is an application of Lemma \ref{P(0+)} relating the asymptotic states of a one-parameter semigroup and its van Hove limit.

\begin{proposition} \label{etat_asymp}
Let $A:\R\rightarrow \text{M}_n(\C)$ be an analytic function, with $A(0) = \sum a_kP_k$ anti-hermitic (we suppose that the $a_k$'s are pairwise different), $\ker A(\eps)\neq \{0\}$ and $\norm{\e^{tA(\eps)}} = 1$. Suppose that
\[
\e^{s\sum P_k A'(0)P_k} \xrightarrow[s\rightarrow\infty]{} Q,
\]
with $\tr Q=1$. Then, there exists an $\eps_0>0$ such that
\[
\e^{tA(\eps)} \xrightarrow[t\rightarrow\infty]{} Q + O(\eps), \quad \forall \eps\in\left]0,\eps_0\right].
\]
\end{proposition}
\begin{proof}
We first fix some notation: write 
\[
\spec A(\eps) = \{a_i(\eps): i\in\{0,\dots,m\}\}, 
\]
with $a_i:\R\rightarrow \C$ continuous for all $i$. Since the null space of $A(\eps)$ is non-trivial, we can suppose that $a_0\equiv 0$. We have the expansion
\[
a_i(\eps) = a_i(0) + \eps^{1/p_i} \lambda_i + O(\eps^{2/p_i}),
\]
where $\lambda_i$ is an eigenvalue of $\sum P_kA'(0)P_k$ and $p_i\in\N$ is the branching order of~$a_i(0)$. 

Recall that  the hypothesis 
\(
\e^{s\sum P_k A'(0)P_k} \xrightarrow[s\rightarrow\infty]{} Q
\)
 is equivalent to  
 \[
 \spec\ \biggl(\sum P_kA'(0)P_k\biggr) \setminus\{0\} \subset \{\lambda\in\C: \re\lambda<0\},
 \]
with $0\in \spec \sum P_kA'(0)P_k$ being
 a semisimple eigenvalue (in fact, simple since $\tr Q=1$) and $Q$ its spectral projection.
Hence, there exists an $\eps_0>0$ such that, except when $i=0$, $\re a_i(\eps) < 0$ for all $0<\eps\leq\eps_0$. This ensures that $\e^{tA(\eps)}$ converges to the spectral projection of $0\in\spec A(\eps)$, which we will call $P(\eps)$. Now, we can make use of Lemma \ref{P(0+)} with $T(\eps)=\e^{A(\eps)}$, obtaining that $P(\eps) = P(0^+) + O(\eps)$ for all  $0<\eps\leq\eps_0$.

Finally, observe that
\begin{align*}
P_0\frac{d}{d\eps}\biggr|_{\eps=0}\e^{A(\eps)}P_0 &= P_0 \frac{d}{d\eps}\biggr|_{\eps=0} \biggl\{ 1 + \eps\int_0^1ds\ \e^{-sA(0)}A'(0)\e^{sA(0)} + O(\eps^2) \biggr\}\e^{A(0)} P_0 \\
	&= P_0 A'(0) P_0.
\end{align*}
Hence, again thanks to Lemma \ref{P(0+)}, $P(0^+)$ is a sub-projection  of $Q$. But $\tr Q = 1$, so that $P(\eps) = Q+O(\eps)$ for all $0<\eps\leq\eps_0$.
\end{proof}

\subsection{Application to Repeated Interaction Systems}

We start with the regime $\lambda\rightarrow 0$.

\begin{theorem} \label{etat_asymp_lambda}
Suppose that $M_S$ is of finite type $\text{I}_n$ and that the effective dynamics $\varphi_\text{eff}^s$ given by Theorem \ref{lambda1} converges towards a projection $P:M_S\rightarrow M_S$ of rank 1. Then, there exists a $\lambda_0>0$ and a $\tau$-periodic family $\omega_\lambda^t\in \mathcal E_S$ such that
\[
\omega\bigl(\varphi_{\text{res}}^t(x)\bigr) - \omega_\lambda^t(x) \xrightarrow[t\rightarrow\infty]{} 0,
\]
for all $0<\lambda\leq\lambda_0$, $\omega\in \mathcal E_S$  and $x\in M_S$. Moreover,
\[
\omega_\lambda^t(x) = \frac{1}{n}\tr \bigl(P\alpha_S^t(x)\bigr) + O(\lambda^2\norm x).
\]
\end{theorem}
\begin{proof}
After the proof of Theorem \ref{lambda1} (whose hypothesis always hold in finite dimension), we can write $T(\lambda,\tau) = \e^{\tau A(\lambda^2)}$ with $A:\left]-\eps_0,\eps_0\right[\rightarrow B(M_S)\cong \text M_{n^2}(\C)$ analytic. Now, a direct application of Proposition \ref{etat_asymp} (recall from Proposition \ref{prop_T} that $\norm{T(\lambda,\tau)} = 1$) gives
\[
T(\lambda,\tau)^k  \xrightarrow[k\rightarrow\infty]{}  P + O(\lambda^2) =: P(\lambda^2).
\]
Since $1\in M_S$ is a fixed point for $T(\lambda,\tau)$, the image of $P(\lambda^2)$ is $\C 1\subseteq M_S$. The result follows with 
\[
\omega_\lambda^t(x) = \frac{1}{n}\tr (P(\lambda^2)\varphi_{SE}^t(x)). 
\]
Observe that $\omega_\lambda^t(x)$ is $\tau$-periodic, for
\[
P(\lambda^2)\varphi_{SE}^\tau(x) = P(\lambda^2)\E_S\varphi_{SE}^\tau(\E_Sx) = P(\lambda^2)x.   
\]
\end{proof}
\begin{remark}
The state
\(
x\mapsto \frac{1}{n}\tr (P\alpha_S^t(x))
\)
is also  $\tau$-periodic sin\-ce  $P$ commutes with $\alpha_S^\tau$ and $PM_S = \C1$.
\end{remark}

Now, we state the result for the regime $\tau\rightarrow 0$, $\lambda^2\tau\rightarrow 0$. We face two extra difficulties:
\begin{enumerate}
\item
	$T(\lambda,\tau)$ cannot be seen as a function of $\eps=\lambda^2\tau$; hence, in order to use Proposition \ref{etat_asymp}, one has to parametrize analytically  the convergences $\tau\rightarrow 0$, $\lambda^2\tau\rightarrow~0$.
\item
	Once we acknowledge the necessity of the previous step, it still has to be shown that one can write $T(\lambda(\eps),\tau(\eps)) = \e^{\tau(\eps)A(\eps)}$, with $A$ analytic.
\end{enumerate}

\begin{lemma} \label{T_bianalyt}
The function $T: (\lambda,\tau)\in \R^2 \mapsto \E_S\e^{\tau(\delta_S+\J\lambda[v,\cdot])}\in B(M_S)\cong \text M_{n^2}(\C)$ can be written, for $\lambda$ and $\tau$ small enough, as
\[
T(\lambda,\tau) = \e^{\tau F(\lambda^2\tau,\tau)}, 
\]
where $F:\R^2\rightarrow B(M_S)$ is analytic in a neighbourhood of $(0,0)$ and $F(0,0) = \delta_S$.
\end{lemma}
\begin{proof}
Indeed, we have the convergent power series expansion
\[ 
\e^{\tau(\delta_S+\J\lambda\tau[v,\cdot])} = \sum_{n=0}^\infty \frac{\tau^n}{n!} \sum_{k=0}^n (\J\lambda\tau)^k  \sum_{\mod\alpha=n-k} \delta_S^{\alpha_0}[v,\cdot]\delta_S^{\alpha_1}\cdots\delta_S^{\alpha_{k-1}}[v,\cdot]\delta_S^{\alpha_k},
\] 
where the multiindex $\alpha$ belongs to $\N^{k+1}$ and $\mod\alpha = \sum_{i=0}^k \alpha_i$. After composing with $\E_S$, the terms with odd $k$ vanish and we get
\begin{gather*}
\E_S\e^{\tau(\delta_S+\J\lambda\tau[v,\cdot])} = 1 + \sum_{n=1}^\infty\sum_{m=0}^{\lfloor n/2 \rfloor} \frac{1}{n!}\tau^{n+m}(\J\lambda^2\tau)^{m} C_{n,m}, \\
C_{n,m} = \E_S\sum_{\mod\alpha=n-2m} \delta_S^{\alpha_0}[v,\cdot]\delta_S^{\alpha_1}\cdots\delta_S^{\alpha_{2m-1}}[v,\cdot]\delta_S^{\alpha_{2m}}.
\end{gather*}
Now, if $\lambda$ and $\tau$ are small enough, the logarithm series
\[
\sum_{N=1}^\infty \frac{(-1)^{N+1}}{N} \biggl(\sum_{n=1}^\infty\sum_{m=0}^{\lfloor n/2 \rfloor} \frac{1}{n!}\tau^{n+m}(\J\lambda^2\tau)^{m} C_{n,m}\biggr)^N
\]
converges and gives the existence of an $F$ which is analytic and satisfies $\e^{\tau F(\lambda^2\tau,\tau)} = T(\lambda,\tau)$. Observe that $C_{1,0} = \delta_S$, so that $F(0,0) = \delta_S$.
\end{proof}

\begin{theorem} \label{etat_asymp_tau}
Let $\lambda(\eps)$ and $\tau(\eps)$, with $\eps\in\R$, be two meromorphic parametrizations of $\lambda$ and $\tau$ such that
\[
\lambda(\eps)^2\tau(\eps) = \eps,\quad  \tau(\eps)\xrightarrow[\eps\rightarrow 0]{} 0.
\]
Suppose that $M_S$ is of finite type $\text{I}_n$ and that the effective dynamics $\varphi_\text{eff}^s$ given by Theorem \ref{tau} converges, as $s\rightarrow\infty$, towards a projection $P\in B(M_S)$ of rank 1. Then, there exists a $\tau(\eps)$-periodic family $\omega_\eps^t\in \mathcal E_S$ and an $\eps_0>0$ such that
\[
\omega\bigl(\varphi_{\text{res}}^t(x)\bigr) - \omega_\eps^t(x) \xrightarrow[t\rightarrow\infty]{} 0,
\] 
for all $\eps\in\left]-\eps_0,\eps_0\right[$, $\omega\in \mathcal E_S$ and  $x\in M_S$. Moreover, 
\[
\omega_\eps^t(x) = \frac{1}{n}\tr (P\alpha_S^t(x)) + O(\eps^2\norm x).
\]
\end{theorem}
\begin{proof}
Let $F(\lambda^2\tau,\tau)$ be the analytic function given by {Lem\-ma}~\ref{T_bianalyt} and consider the family of one-parameter groups
\[
t\mapsto \e^{tA(\eps)},\quad A(\eps) = F\bigl( \lambda(\eps)^2\tau(\eps),\tau(\eps) \bigr) = F\bigl(\eps,\tau(\eps)\bigr).
\]
Observe that $A(\eps)$ is analytic. In order to relate $\e^{tA(\eps)}$ and $\varphi_\text{eff}^s$, fix $s>0$ and let $m(\eps) = \lfloor s/(\lambda(\eps)\tau(\eps))^2\rfloor$, so that 
\[
s/(\lambda(\eps)^2\tau(\eps)) = m(\eps)\tau(\eps) + t_1(\eps),\quad 0\leq t_1(\eps)<\tau(\eps). 
\]
From now on, we will drop the dependence in $\eps$ of $m, \lambda, \tau$ and $t_1$ (this should cause no confusion). 
Write
\[ 
\Norm{ T(\lambda,\tau)^m \alpha_S^{-m\tau} - \varphi_\text{eff}^s }  = \Norm{ \varphi_\text{res}^{s/(\lambda^2\tau)}\alpha_S^{-s/(\lambda^2\tau)} - \varphi_\text{eff}^s } + \Norm{T(\lambda,\tau)^m \alpha_S^{-m\tau} - \varphi_\text{res}^{s/(\lambda^2\tau)}\alpha_S^{-s/(\lambda^2\tau)}}.
\] 
As $\eps\rightarrow 0$, the first term vanishes by Theorem \ref{tau} and the fact that we are dealing with (finite) matrices. The second equals
\[
\Norm{\alpha_S^{t_1} - \E_S\varphi_{SE}^{t_1}} \xrightarrow[\eps\rightarrow 0]{} 0.
\]
Since $T(\lambda,\tau)^m \alpha_S^{-m\tau} =  \e^{m\tau A(\eps)}\e^{-m\tau A(0)}$ and
\begin{align*}
\Norm{ \e^{sA(\eps)/\eps}\e^{-sA(0)/\eps} - \e^{m\tau A(\eps)}\e^{-m\tau A(0)} } 
	&=\Norm{ \e^{m\tau A(\eps)} (\e^{t_1 A(\eps)}\e^{-t_1 A(\eps)} - 1) \e^{-m\tau A(0)} } \\
	&\leq \e^{s\sup_{\eps\leq\eps_0}\norm{A(\eps)}} \Norm{ \e^{t_1 A(\eps)}\e^{-t_1 A(\eps)} - 1 } \xrightarrow[\eps\rightarrow 0]{} 0,
\end{align*}
what we get is that
\[
\varphi_\text{eff}^s = \lim_{\eps\rightarrow 0} \e^{sA(\eps)/\eps} \e^{-sA(0)/\eps}.
\]
By uniqueness of both limits and generators of semigroups, we see that $-\frac{1}{2}(\E_S[v,\cdot]^2)^\natural = A'(0)^\natural$. Hence, we can apply Proposition \ref{etat_asymp} to conclude that there exists an $\eps_0>0$ such that
\[
T\bigl( \lambda(\eps),\tau(\eps) \bigr)^k \xrightarrow[k\rightarrow\infty]{} P + O(\eps^2) =: P(\eps^2),
\]
for all $\eps\in \left]-\eps_0,\eps_0\right[$. The proof ends in the same way as that of Theorem \ref{etat_asymp_lambda}.
\end{proof}

\begin{remark}
Since we ask from $\tau(\eps)$ to be analytic around 0, we can as well just assume that $\tau(\eps) = \eps^n$, with $n\geq 1$. Now, the restriction $\lambda(\eps)^2\tau(\eps)=\eps$ on the parametrizations of $\lambda$ and $\tau$---which seems to be essential in our approach---implies that
\[
\lambda(\eps) = \eps^{(1-n)/2}
\]
(further restricting $n$ to be odd),  showing that our theorem cannot say anything of a regime in which both $\lambda$ and $\tau$ go to zero.
\end{remark}

\begin{remark}
Let $\tilde\eps \in \left]-\eps_0,\eps_0\right[$. The convergence 
\[
\omega\bigl(\varphi_{\text{res}}^{t+n\tau}(x)\bigr)\xrightarrow[n\rightarrow\infty]{} \omega_{\tilde\eps}^t(x)
\]
shows that $\omega_{\tilde\eps}$ depends on the values $\lambda(\tilde\eps)$ and $\tau(\tilde\eps)$, but does not depend on the choice of parametrizations.
\end{remark}

This last remark suggests that Theorem \ref{etat_asymp_tau} would be better stated without any reference to the parametrizations. To this effect, we could consider the set
\[
\bigcup_{\substack{\text{admissible} \\ \text{parametrizations}}} \bigl\{ (\lambda(\eps),\tau(\eps)): \eps \in \left]-\eps_0, \eps_0\right[ \bigr\}, 
\]
where $\eps_0>0$ depends on the parametrization. However, we lack any description ot this set which does not actually mention the parametrizations; this is the reason why we prefer to state Theorem \ref{etat_asymp_tau} as we did.

\section{A concrete example} \label{example}

In the simplest instance of a repeated interaction system, both the small system and the chain element are spins. This case falls under the hypothesis of \cite{AJ}, in which the effective dynamics for the regime $\lambda\rightarrow 0$ is explicitely calculated (for some specific choice of the interaction). Also, in \cite{BJM}, explicit conditions for the existence of an asymptotic time-periodic state are found, and the asymptotic state itself is computed at zero-th order in $\lambda^2$. Here, we illustrate how this last result can be recovered as an application of Theorem \ref{etat_asymp_lambda}. 

Let us specify the model. We choose the representation
\[
M_S = M_E = \text{M}_2(\C),\quad \H_S=\H_E = \C^2,
\]
and suppose that the free evolution of observables is given by the hamiltonians
\[
h_S = \begin{pmatrix} 0 & 0 \\ 0 & S \end{pmatrix} \in M_S,\quad h_S = \begin{pmatrix} 0 & 0 \\ 0 & E \end{pmatrix} \in M_E.
\]
As for the interaction, we take
\[
v = \begin{pmatrix} 0 & 1 \\ 0 & 0 \end{pmatrix} \otimes \begin{pmatrix} a & b \\ c & d \end{pmatrix} +  \begin{pmatrix} 0 & 0 \\ 1 & 0 \end{pmatrix} \otimes \begin{pmatrix} \bar a & \bar c \\ \bar b & \bar d \end{pmatrix} \in  M_S\otimes M_E.
\]
Finally, we assume that the chain is initially in thermal equilibrium at inverse temperature $\beta$; that is, 
\[
\omega_E \begin{pmatrix} x_{00} & x_{01} \\ x_{10} & x_{11} \end{pmatrix} = \frac{x_{00} + x_{11}\e^{-\beta E}}{1+\e^{-\beta E}}. 
\]

To make calculations, let $\{\epsilon_0,\epsilon_1\}$ be the canonical basis of $\C^2$ and consider the basis of $\text{M}_2(\C)$ given by $u_{kl} = \ket{\epsilon_k}\bra{\epsilon_l}$, with $k,l,\in \{0,1\}$.
We find that
\begin{align*}
\alpha_S^t(u_{00}) &= u_{00},&  \alpha_S^t(u_{01}) &= \e^{\J tS} u_{01}, \\
\alpha_S^t(u_{10}) &= \e^{-\J tS} u_{10},&  \alpha_S^t(u_{11}) &= u_{11},
\end{align*}
so that assuming that $S\neq 0$ and that $\e^{\J\tau S}\neq \e^{-\J\tau S}$, the spectral averaging in the formula for the generator of the effective dynamics $\varphi_\text{eff}^s$ must be taken with respect to the projections
\[
P_0 = P_{00}+P_{11},\quad P_+ = P_{01},\quad P_- = P_{10},
\]
where $P_{kl} = \tr\bigl(u_{kl}^*(\cdot)\bigr)u_{kl}$.  Observe that, if $\tau$ is small enough, $\e^{\J\tau S}\neq \e^{-\J\tau S}$.

Since we are interested in the asymptotic state of the effective dynamics when $\lambda\rightarrow 0$, we must compute the spectral projection of the kernel of
\begin{align*}
\delta_\text{eff} &:= -(\E_S\varphi_{SE,2}^\tau)^\natural \\ &=  -P_0 \E_S\varphi_{SE,2}^\tau P_0 -  P_- \E_S\varphi_{SE,2}^\tau P_- - P_+ \E_S\varphi_{SE,2}^\tau P_+.
\end{align*}
Now, if $\bra{u_{01}} \delta_\text{eff} \ket{u_{01}}$ and $\bra{u_{10}} \delta_\text{eff} \ket{u_{10}}$ do not vanish, that spectral projection is, essentially, the one of $P_0 \delta_\text{eff} |_{P_0 M_2(\C)}$. Identifying $P_0 M_2(\C) \cong \C^2$ through the basis $\{u_{00}, u_{11}\}$, this operator is the  $2\times 2$ matrix
\[
\begin{pmatrix} \bra{u_{00}}\delta_\text{eff}\ket{u_{00}} & \bra{u_{00}}\delta_\text{eff}\ket{u_{11}} \\ \bra{u_{11}}\delta_\text{eff}\ket{u_{00}} & \bra{u_{11}}\delta_\text{eff}\ket{u_{11}} \end{pmatrix}.
\]
But $0 = \delta_\text{eff}(1) = \delta_\text{eff}(u_{00}+u_{11})$,  so that this matrix has the form $\bigl( \begin{smallmatrix} \delta_0 & -\delta_0 \\ -\delta_1 & \delta_1\end{smallmatrix} \bigr)$, with 
\[
\delta_0=\bra{u_{00}}\delta_\text{eff}\ket{u_{00}}, \quad\delta_1=\bra{u_{11}}\delta_\text{eff}\ket{u_{11}}. 
\]
The spectral projection of its kernel is 
\[
Q  = \frac{1}{\delta_0+\delta_1} \begin{pmatrix} \delta_1 & \delta_0 \\ \delta_1 & \delta_0 \end{pmatrix},
\]
and we find that
\begin{align*}
\delta_0 &= \frac{-2}{1+\e^{-\beta E}} \left\{ \e^{-\beta E}|b|^2 \frac{1-\cos\tau(E-S)}{(E-S)^2} + |c|^2 \frac{1-\cos\tau(E+S)}{(E+S)^2} \right\}, \\
\delta_1 &= \frac{-2}{1+\e^{-\beta E}} \left\{ |b|^2 \frac{1-\cos\tau(E-S)}{(E-S)^2} + \e^{-\beta E}|c|^2 \frac{1-\cos\tau(E+S)}{(E+S)^2} \right\}.
\end{align*}

We are in a position to compute the asymptotic state of the weak limit. As Theorem \ref{etat_asymp_lambda} ensures, it coincides at order zero with the one of the restricted dynamics, computed in \cite{BJM}. As sufficient conditions for its existence we recover also the result in \cite{BJM}.
\begin{proposition}
Suppose that  $S\neq 0$  and $|b|^2+|c|^2\neq 0$, and let $\varphi_\text{eff}^s$ be the effective dynamics given by Theorem \ref{lambda1}. There exists some $\tau_0>0$ such that
\[
\omega(\varphi_\text{eff}^s(x)) \underset{s\rightarrow\infty}{\longrightarrow} \frac{1}{\delta_0+\delta_1} \tr \left( \begin{pmatrix} \delta_1 & 0 \\ 0 & \delta_0 \end{pmatrix} x \right),
\]
for all $\omega \in \mathcal E_S$, $x\in M_S$ and $\tau\leq \tau_0$.
\end{proposition}
\begin{proof}
Let $x= \bigl( \begin{smallmatrix} x_{00} & x_{01} \\ x_{10} & x_{11} \end{smallmatrix} \bigr)\in M_S$. The computations above show that---provided there is convergence---
\[
\e^{s\delta_\text{eff}}(x) \underset{s\rightarrow\infty}{\longrightarrow} \frac{\delta_1 x_{00} + \delta_0 x_{11}}{\delta_0+\delta_1} \begin{pmatrix} 1 & 0 \\ 0 & 1 \end{pmatrix}.
\]
It remains to see, for every small enough $\tau$, that there is indeed convergence.

With respect to the basis $\{ u_{00},u_{11},u_{01},u_{10} \}$, 
\[
\delta_\text{eff} = \begin{pmatrix} \delta_0 & -\delta_0 & 0 & 0 \\ -\delta_1 & \delta_1 & 0 & 0 \\ 0 & 0 & \bra{u_{01}}\delta_\text{eff}\ket{u_{01}} & 0 \\ 0 & 0 & 0 & \bra{u_{10}}\delta_\text{eff}\ket{u_{10}} \end{pmatrix}.
\]
The eigenvalues of this matrix are
\[
0,\ \delta_0+\delta_1,\ \bra{u_{01}}\delta_\text{eff}\ket{u_{01}} \text{ and } \bra{u_{10}}\delta_\text{eff}\ket{u_{10}}, 
\]
and we have to check that, except for 0, their real part is strictly negative. Since $|b|^2 + |c|^2\neq 0$, one has that  $\delta_0+\delta_1 = \re (\delta_0+\delta_1) < 0$. As for the others, up to order $\tau^2$ we have that 
\begin{align*}
\re \bra{u_{01}}\delta_\text{eff}\ket{u_{01}}
&= \frac{-\tau^2}{2(1+\e^{-\beta E})} \left\{ \begin{pmatrix} \bar a & \bar c \\ \bar b & \bar d \end{pmatrix} \begin{pmatrix} -c & a-d \\ 0 & c \end{pmatrix} - \begin{pmatrix} -\bar b & \bar a - \bar d \\ 0 & \bar b \end{pmatrix} \begin{pmatrix} a & b \\ c & d \end{pmatrix} \right. \\
	&\qquad\left. + \e^{-\beta E} \begin{pmatrix} a & b \\ c & d \end{pmatrix} \begin{pmatrix} -\bar b & \bar a - \bar d \\ 0 & \bar b \end{pmatrix} - \e^{-\beta E} \begin{pmatrix} -c & a-d \\ 0 & c \end{pmatrix} \begin{pmatrix} \bar a & \bar c \\ \bar b & \bar d \end{pmatrix} \right\}_{01} \\
	&= \frac{-\tau^2}{2(1+\e^{-\beta E})} \bigl\{ |a|^2 + |b|^2 + |c|^2 + |d|^2 - 2\bar ad \\ &\qquad + \e^{-\beta E}\bigl( |a|^2 + |b|^2 + |c|^2 + |d|^2 - 2a\bar d \bigr) \bigr\} \\
	&\leq \frac{-\tau^2}{2} \bigl( |b|^2 + |c|^2 \bigr) < 0,
\end{align*}
whereas 
\begin{align*}
\re \bra{u_{10}}\delta_\text{eff}\ket{u_{10}}
&= \frac{-\tau^2}{2(1+\e^{-\beta E})} \left\{ \begin{pmatrix} \bar a & \bar c \\ \bar b & \bar d \end{pmatrix} \begin{pmatrix} c & 0 \\ d-a & -c \end{pmatrix} - \begin{pmatrix} \bar b & 0 \\ \bar d-\bar a & -\bar b \end{pmatrix} \begin{pmatrix} a & b \\ c & d \end{pmatrix} \right. \\
	&\qquad\left. + \e^{-\beta E} \begin{pmatrix} a & b \\ c & d \end{pmatrix} \begin{pmatrix} \bar b & 0 \\ \bar d-\bar a & -\bar b \end{pmatrix} - \e^{-\beta E} \begin{pmatrix} c & 0 \\ d-a & -c \end{pmatrix} \begin{pmatrix} \bar a & \bar c \\ \bar b & \bar d \end{pmatrix} \right\}_{10} \\
	&= \frac{-\tau^2}{2(1+\e^{-\beta E})} \bigl\{ |a|^2 + |b|^2 + |c|^2 + |d|^2 - 2a\bar d \\ &\qquad + \e^{-\beta E}\bigl( |a|^2 + |b|^2 + |c|^2 + |d|^2 - 2\bar ad \bigr) \bigr\} \\
	&\leq \frac{-\tau^2}{2} \bigl( |b|^2 + |c|^2 \bigr) < 0. 
\end{align*}
\end{proof}

\appendix 
\section{The Dyson series} \label{appDyson}

In this appendix we collect the results we need on the perturbation series known as the \emph{Dyson series}. Proofs can be found in \cite{bratteli-robinson}, for example.

\begin{theorem} \label{dyson}
Let $X$ be a Banach space with predual $X_*$ and  
\[
A_0:\dom A_0 \subseteq X\rightarrow X
\] 
the generator of a $*$-weakly-continuous semigroup $\{S^t\}_{t\in\R_+}$. Consider the perturbation $A(\eps) = A_0 + \eps A_1$, where $A_1\in B(X)$. We have that $A(\eps)$ generates a $*$-weakly-continuous semigroup too, which we will denote by $\{S(\eps)^t\}_{t\in\R_+}$. It satisfies
\[
S(\eps)^t = S^t + \sum_{n\geq 1} \eps^n\int_0^t dt_n\cdots\int_0^{t_2}dt_1\ S^{t_1} A_1 S^{t_2-t_1}A_1 \cdots A_1 S^{t_n-t_{n-1}} A_1 S^{t-t_n}.
\]
Here, the integrals are defined pointwise in the weak-$*$ topology and give a convergent series for every $\eps>0$.
\end{theorem}

\begin{remark} \label{err_dyson}
Given any one-parameter $*$-weakly-continuous semigroup, there always exist constants $M\geq 1$ and $\beta\geq 0$ such that
$\|S^t\| \leq M\e^{\beta t}$ 
(see \cite[Proposition 3.1.3]{bratteli-robinson}). Hence, the $n$-th term
\[
S_n^t = \int_0^t dt_n\cdots\int_0^{t_2}dt_1\ S^{t_1} A_1 S^{t_2-t_1}A_1 \cdots A_1 S^{t_n-t_{n-1}} A_1 S^{t-t_n}
\]
in the Dyson series satisfies
\[
\norm{S_n^t} \leq \frac{t^n}{n!}M^{n+1}\e^{\beta t}\norm{A_1}^n.
\]
Therefore, the error after adding up the first $n-1$ terms is bounded by
\begin{equation} \label{f_n}
\eps^{n}t^{n}\e^{\beta t}\sum_{k\geq n}\frac{(\eps t)^{k-n}}{k!}M^{k+1}\norm{A_1}^k =: \e^{\beta t}f_n(\eps t) \eps^nt^n,
\end{equation}
where $f_n:\R_+\rightarrow\R_+$ is a continuous and increasing function.
\end{remark}

\section*{acknowledgements}
I am sincerely thankful to professor Alain Joye for his guidance and strong support. This article was written while visiting the mathematical engineering department of Universidad de Chile, where it felt (not surprisingly) like home. Special thanks to professor Alejandro Maass for making that possible.

\bibliographystyle{spmpsci}    
\bibliography{/biblio.bib}

\end{document}